\documentclass[proof]{pasj00}
\draft 

\begin{document}


\SetRunningHead{Goto et al.}{Running Head}
\Received{2002 April 9}
\Accepted{2002 May 15}
\title{Composite Luminosity Functions based on the Sloan Digital Sky Survey ``Cut and Enhance'' Galaxy Cluster Catalog}


 \author{%
   Tomotsugu \textsc{Goto}\altaffilmark{1,2,3},
   Sadanori \textsc{Okamura}\altaffilmark{4},
   Timothy A. \textsc{Mckay}\altaffilmark{5},\\
   Neta A. \textsc{Bahcall}\altaffilmark{6},
   James \textsc{Annis}\altaffilmark{7},
   Mariangela \textsc{Bernardi}\altaffilmark{2},\\
   J. \textsc{Brinkmann}\altaffilmark{8},
   Percy L. \textsc{G{\' o}mez}\altaffilmark{2},
   Sarah \textsc{Hansen}\altaffilmark{5},\\
   Rita S. J. \textsc{Kim}\altaffilmark{9},
   Maki \textsc{Sekiguchi}\altaffilmark{1} 
   and Ravi K. \textsc{Sheth}\altaffilmark{10}
}   
 \altaffiltext{1}{Institute for Cosmic Ray Research, University of
   Tokyo,\\ Kashiwanoha, Kashiwa, Chiba 277-0882, Japan}
 \altaffiltext{2}{Department of Physics, Carnegie Mellon University, \\
 5000 Forbes Avenue, Pittsburgh, PA 15213-3890, USA}
 \altaffiltext{3}{tomo@cmu.edu}
 \altaffiltext{4}{Department of Astronomy and Research Center for the
   Early Universe,\\ School of Science, University of Tokyo, Tokyo
   113-0033, Japan}
 \altaffiltext{5}{University of Michigan, Department of Physics,\\ 500
   East University, Ann Arbor, MI 48109, USA}
 \altaffiltext{6}{Princeton University Observatory, Princeton, NJ 08544,   USA}
 \altaffiltext{7}{Fermi National Accelerator Laboratory, P.O. Box 500,
   Batavia, IL 60510, USA }
 \altaffiltext{8}{Apache Point Observatory,\\
     2001 Apache Point Road,
     P.O. Box 59, Sunspot, NM 88349-0059, USA}
 \altaffiltext{9}{Department of Physics and Astronomy, The Johns Hopkins
   University,\\ 3400 North Charles Street, Baltimore, MD 21218-2686, USA}
 \altaffiltext{10}{ Department of Physics and Astronomy
 University of Pittsburgh\\
 3941 O'Hara Street
 Pittsburgh, PA 15260  }


%


\maketitle

\begin{abstract}

 We present here results on the composite luminosity functions (LF)
  of galaxies in the clusters of galaxies selected from the Cut
 and Enhance cluster catalog (CE) of the Sloan Digital Sky
 Survey.
 We constructed composite LFs in the five SDSS bands,
 $u,g,r,i$ and $z$, using 204 CE clusters ranging from
 $z$ = 0.02 to $z$ = 0.25 . 
 Background and foreground galaxies were subtracted from the
 LF using an annular region around clusters to take
 large-scale, galaxy-number-count variances into consideration.  
 A LF of each cluster  was weighted according to the richness and number of
 contributing galaxies to construct the composite LF.  
 Taking advantage of accurate photometry of SDSS,
 we used photometric redshifts to construct composite luminosity
 functions and thus study a large number of clusters.
 The robustness of the weighting scheme was tested using a Monte-Carlo simulation.  
  The best-fit Schechter parameters are
 ($M^*$,$\alpha$)=($-$21.61$\pm$0.26, $-$1.40$\pm$0.11),($-$22.01$\pm$0.11,
 $-$1.00$\pm$0.06),($-$22.21$\pm$0.05, $-$0.85$\pm$0.03), ($-$22.31$\pm$0.08,
 $-$0.70$\pm$0.05) and ($-$21.36$\pm$0.06, $-$0.58$\pm$0.04) in  $u, g,r,i$ and
 $z$, respectively.
 We find that the slope of composite LFs becomes flatter toward a redder
 color band. Compared with the field LFs of the SDSS,
 the cluster LFs have brighter characteristic magnitude and flatter slopes
 in the $g, r, i$ and $z$ bands.  
 These results are consistent with the hypothesis that the cluster LF has two
 distinct underlying populations i.e. the bright end of the LF is
 dominated by bright early types that follow a Gaussian-like luminosity
 distribution, while the faint-end of the cluster LF is a steep power-law
 like function dominated by star-forming (bluer) galaxies. 
 We also studied the composite LFs for early-type and
 late-type galaxies using profile fits, a concentration parameter and $u-r$
 color to classify galaxy morphology.
 A strong dependence of LF on galaxy morphology was found.
 The faint end slope of the LF is always flatter for early-type
 galaxies than late-type, regardless of the passband and methodology.
 These results are consistent with the hypothesis that the cluster
 regions are dominated by bright elliptical galaxies.
 This work also provides a good low-redshift benchmark for
 on-going multi-color photometric studies of high redshift clusters of
 galaxies using 4--8 m class telescopes. 

\end{abstract}

\section{Introduction}
\label{intro}

 The luminosity function (LF) of galaxies within clusters of
galaxies is a key tool for understanding the role of the environment on galaxy
formation and evolution.  The shape of the cluster LF as a function of
the galaxy colors and morphologies, as well as a function of the cluster radius or local density, can provide strong observational constraints
on the theories of galaxy formation. For example, Springel et al. (2001)
recently showed that semi--analytical models of hierarchical structure formation could
now explain both the shape of the composite cluster LF ($B$--band LF of Trentham 1998) and the morphology--radius relationship of Whitmore et al. (1993)
using just a simple prescription for the properties of galaxies in clusters
based on their merger and cooling rates (see also Okamoto, Nagashima
2001). Empirically, there is also growing evidence for a correlation between
the shape of the cluster LF and the underlying cluster properties.  Phillipps
et al. (1998) and Driver et al. (1998) show that more
evolved clusters, based on either their density profile or the presence of a
cD galaxy, have flatter faint-end slopes, which they attribute to the
disruption of faint galaxies in the cores of such evolved systems [see the
earlier theoretical work on galaxy cannibalism by Hausman \& Ostriker
(1978)]. In summary, the LF of galaxies in clusters as a function of both the
galaxy and cluster properties is a powerful observational test for theories
of galaxy formation and evolution. The reader is referred to the seminal
review by Binggeli, Sandage, and Tammann (1988), which is still relevant today.

In this paper, we present an analysis of the composite cluster LF
based on the commissioning data of the Sloan Digital Sky Survey (SDSS; see
Gunn et al. 1998; York et al. 2000; Stoughton et al. 2002). This analysis has several key
advantages over previous studies of the composite cluster LF, including
accurate multi-color CCD photometry for all galaxies (in optical passbands
$u,g,r,i$ and $z$; Fukugita et al. 1996), large aerial
 coverage, thus enabling us to make a local correction for the projected field
LF, and finally, the availability of several objectively-measured galaxy
properties like morphology.  Furthermore, we selected our clusters from
the SDSS Cut and Enhance (CE) cluster catalog of Goto et al. (2002), which has
two major benefits over previous cluster samples used for LF studies. First,
the CE catalog was objectively constructed using the latest cluster-finding
algorithms, and therefore has a well-determined selection function (see Goto
et al. 2002). Secondly, CE has obtained an accurate photometric redshift for
each cluster based on the observed color of the E/S0 ridge-line using
the method of J. Annis et al. (in preparation). The error on
this cluster photometric redshift is only $\delta z=0.015$ for $z<0.3$
clusters (see figure. 14 of Goto et al. 2002) and, as we show herein, is
accurate enough to allow us to determine the composite LF for a large sample
of CE clusters without the need for spectroscopic redshifts. Thus, our analysis
of the composite cluster LF is based on one of the largest sample of
clusters  to date.

We present this work now to provide a low-redshift benchmark for
on-going multi-color photometric studies of high redshift clusters of
galaxies.  With the advent of large-area CCD imagers on large telescopes, the
number of distant clusters with such data will increase rapidly over the next
few years; e.g. Kodama et al. (2001) recently presented large-area
multi-color CCD photometry for the distant cluster A 851 ($z=0.41$) using
Suprime-Cam on the Subaru Telescope. Gladders and Yee. (2000) searched distant clusters over 100 deg$^2$ of CCD data.  This paper is organized as follows: In
section 2 we describe the methods used to construct the composite LF of CE
clusters and show our results as a function of passband and morphology. In
section 3 we test the robustness of the analysis, and in section 4 we
summarize our work. Throughout this
paper, we use $h_0$ = 0.7, $\Omega_{\rm{M}}$ = 0.3 and $\Omega_{\Lambda}$ = 0.7.

\section{SDSS Data}

In this section, we outline the data used in this paper. The photometric data
used herein was taken from the SDSS commissioning data, as discussed by York et
al. (2000). Our analysis focuses on the 150 deg$^2$ contiguous area made
 up
from the overlap of SDSS photometric runs 752 \& 756  i.e. \timeform{145D.1}$<$R.A.$<$\timeform{236D.1} and \timeform{-1D.25}$<$Dec.$<$ \timeform{+1D.25}. This is a subset
of the SDSS Early Data Release, as discussed in Stoughton et al. (2002), and
similar to the data used by Scranton et al. (2002) for studying the angular clustering of SDSS galaxies. This photometric data
 reach 5$\sigma$ detection limits for point sources of 22.3, 23.3, 23.1,
 22.3 and 20.8 mag in the $u,g,r,i$ and
 $z$ passbands, respectively (for an airmass of 1.4 and $1''$
seeing)\footnote{The photometry obtained at
this early stage of SDSS is denoted $u^*,g^*,r^*,i^*$, and $z^*$ to stress the
preliminary nature of the calibration.}.
 The photometric uniformity of the data across the whole area is less
than 3\% [see Hogg et al. (2001) and Smith et al. (2002) for photometric calibration],
 while the star-galaxy separation is robust to $r\simeq$21.0 (Scranton
 et al. 2002). This is
 significantly better than previous photographic surveys, which suffer from
 larger plate-to-plate photometric fluctuations and a lower dynamic range
 [see Lumsden et al. 1997 for the problems associated with photographic studies
of the cluster composite LF].  For each galaxy, we used the model
magnitude computed by the PHOTO data analysis pipeline, which has been shown
by Lupton et al. (2001) and Stoughton et al. (2002) to be the optimal
magnitude for faint SDSS galaxies. It is also close to the total magnitude for
the fainter SDSS galaxies.  For a full discussion of the photometric data, and
the galaxy parameters derived from that data, we refer the reader to Lupton et
al. (2001) and Stoughton et al. (2002).

The clusters used herein were drawn from the large sample of CE clusters
presented in Goto et al. (2002), which were selected over the same photometric
runs of 752 \& 756. We only selected the richer systems which were determined
 by the number of galaxies brighter than $-$18th magnitude, ($N_{-18}$).
 The CE clusters used here satisfy the following conditions:

1) Number of galaxies brighter than $-$18th magnitude in the $r$ band ($N_{-18}$) $>$ 20,

2) 0.02$<z<$0.25.

  Condition 1 was used to select richer systems. $N_{-18}$ is defined
as the number of galaxies brighter than $-$18th magnitude in the $r$ band
after subtracting the background using the method described in section
\ref{method} to construct composite LFs. Galaxies within 0.75 Mpc from
a cluster center were used.
 Condition 1 was used to avoid letting small groups with only a few very bright
 galaxies dominate the composite LFs in the weighting scheme. (The
 weighting scheme is explained in detail in section \ref{method}.) Even
 though we used $N_{-18}>$20 as a criteria to select our clusters, we
 show in section \ref{discussion} that our composite LFs were not
 affected by this richness criteria. Since the high redshift clusters ($z\sim$0.3)
 are not imaged to the fainter galaxies, we restricted our clusters to be
 in the range  0.02$<z<$0.25. In total, 204 clusters satisfy these criteria.
 
\section{Analysis and Results}
\label{method}

\subsection{Construction of the Composite Cluster LF}

 We discuss here the construction of the composite luminosity function of
 galaxies within the subsample of the CE clusters discussed above. The first
 critical step in such an analysis is the subtracting the background and
 foreground contamination.  Ideally, one would wish to do this via
 spectroscopic observations, but since the CE cluster catalog contains
 $\sim$2000 galaxies in the region used, it is not feasible to observe
 all clusters spectroscopically.
 Therefore, we must make a statistical correction based on the
 expected contamination from projected field galaxies. One of the main
 advantages of the SDSS data is that such a correction can be estimated
 locally (i.e. free from any galaxy number count variances due to the large-scale
 structure)
 for each cluster since we possess all the photometric data, to the same depth
 and in the same filter set, well outside of the cluster. 
 Indeed, such local background subtraction was thought to be ideal in
 previous works, but was not possible due to the small coverage of the sky.

 For the composite cluster LF, we only used galaxies within 0.75 Mpc of the
 cluster centroid. This radius was 
 determined empirically so as not to lose
 statistics by using a too small radius,
 and not to lose the contrast of clusters
 against the background by using a too large radius.
 Foreground and background contamination were corrected for
 using an annulus around each cluster with an inner radius of 1.5 Mpc and an
 outer radius of 1.68 Mpc. These radii represent a compromise between having as
 large an aperture as possible to avoid removing legitimate cluster galaxies,
 while still providing an accurate estimate of the local projected field
 population.  Since the background/foreground galaxies are themselves highly
 clustered, it is important to obtain as local an estimate as possible.  The
 photometric redshift of each cluster was used to convert these metric
 apertures into angular apertures.  The center of each cluster was taken from
 the CE catalog, and estimated from the position of the peak in the
 enhanced density map of Goto et al. (2002). The cluster centroids were expected
 to be determined with an accuracy better than $\sim$40 arcsec through Monte-Carlo simulation.  When an annulus touches the boundary
of the SDSS data, we corrected for contamination using the number-magnitude
relationship of the whole data set instead (this only affected a few of the
clusters used herein).
 
 Since each sample cluster has a different redshift, each cluster
 reaches the SDSS apparent magnitude limit at different absolute magnitudes. Also, because
 they have various richnesses, the number of galaxies in each cluster
 is different.  To take these different degrees of completeness into account,
 we followed the methodology of Colless (1989) to construct the composite cluster LF.
 The individual cluster LFs are weighted according to the cluster
 richness and the number of clusters which contribute to a given bin. This is
 written as

 \begin{equation} 
    N_{cj}=\frac{m_j}{N_{c0}}\sum_{i}\frac{N_{ij}}{N_{i0}},   
 \end{equation}  

\noindent where $N_{cj}$ is the number of galaxies in the $j$th bin of the
 composite LF, $N_{ij}$ is the number in the $j$th bin of the $i$th cluster LF,
 $N_{i0}$ is the normalization of the $i$th cluster LF, and was measured to be
 the field-corrected number of galaxies brighter than $M_{r^*}=-18$,
 $m_j$ is the number of clusters contributing the $j$th bin and, finally,
 $N_{c0}=\sum_{i}N_{i0}$.  The formal errors on the composite LF were computed
 using

 \begin{equation} 
    \delta N_{cj}=\frac{N_{c0}}{m_j}   \left[\sum_{i} \left(\frac{\delta N_{ij}}{N_{i0}}\right)^2 \right]^{1/2}   ,
 \end{equation} 

 \noindent where $\delta N_{cj}$ and $\delta N_{ij}$ are the errors on the $j$th bin for
 the composite and $i$th cluster, respectively.  In this way we can take
 into account the different degrees of completeness.

 Like other authors, we discarded the Brightest Cluster Galaxy (BCG)
 within 0.75 Mpc of the cluster centroid when constructing the composite
 LF, since such BCGs tend not to follow the cluster LF.  We only used SDSS galaxies
 brighter than $r^*$=21.0, since this is the limit of the SDSS star--galaxy
 separation (Scranton et al. 2002; Lupton et al. 2001). This
 magnitude limit and weighting scheme combined with our cosmology enabled us to dig LF down to $M_{r^*}$=$-$17.5 .  When converting apparent to absolute magnitudes, we assumed a k-correction for the early-type galaxy given by Fukugita et al. (1995).

\begin{figure}[h]
\begin{center}
\includegraphics[scale=0.5]{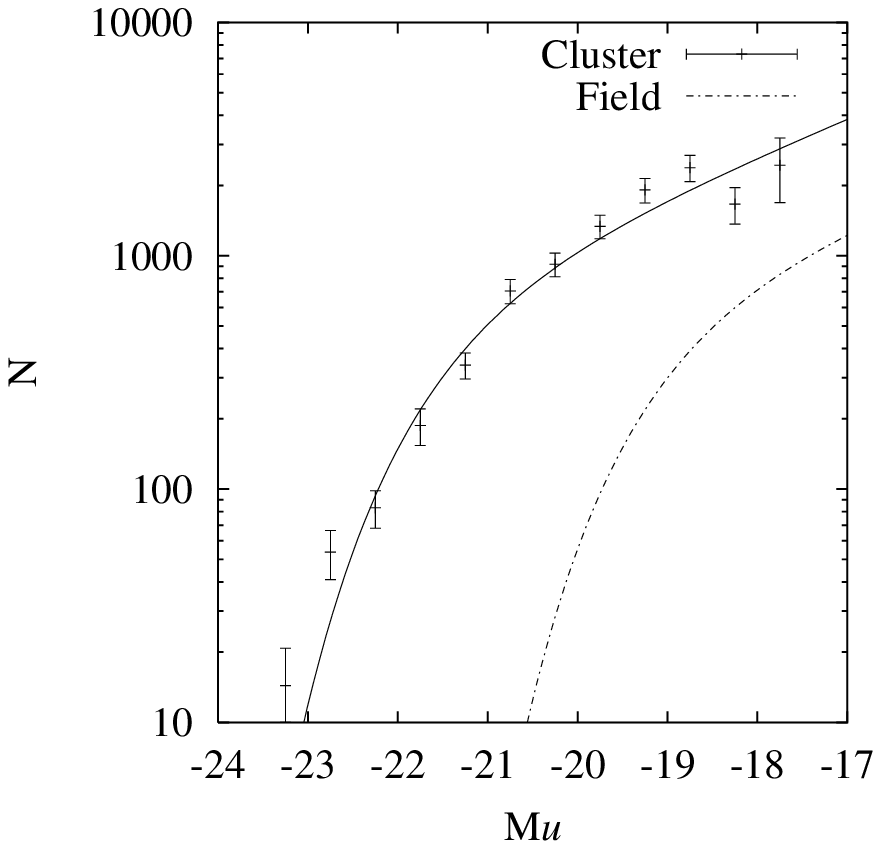}
\includegraphics[scale=0.5]{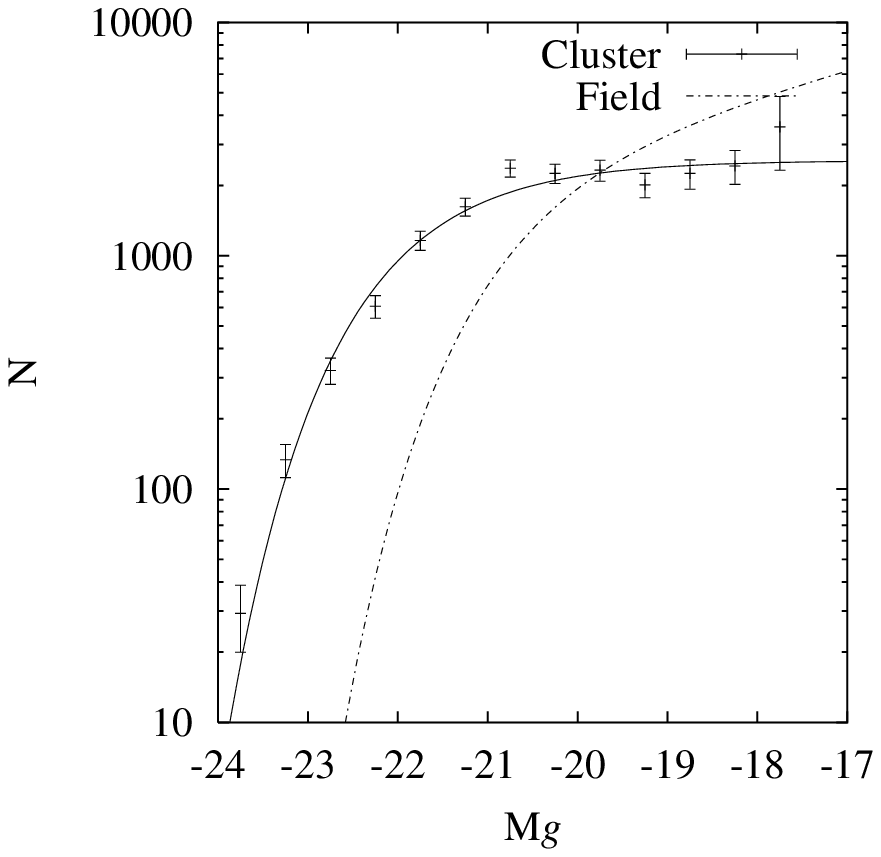}
\includegraphics[scale=0.5]{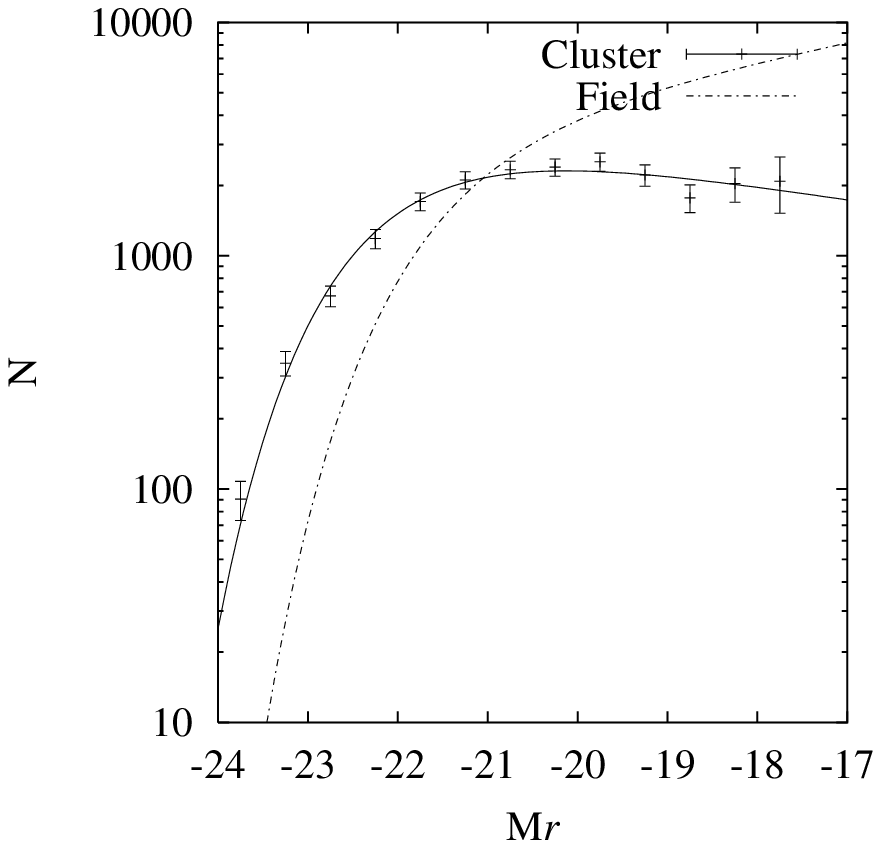}
\end{center}
\begin{center}
\includegraphics[scale=0.5]{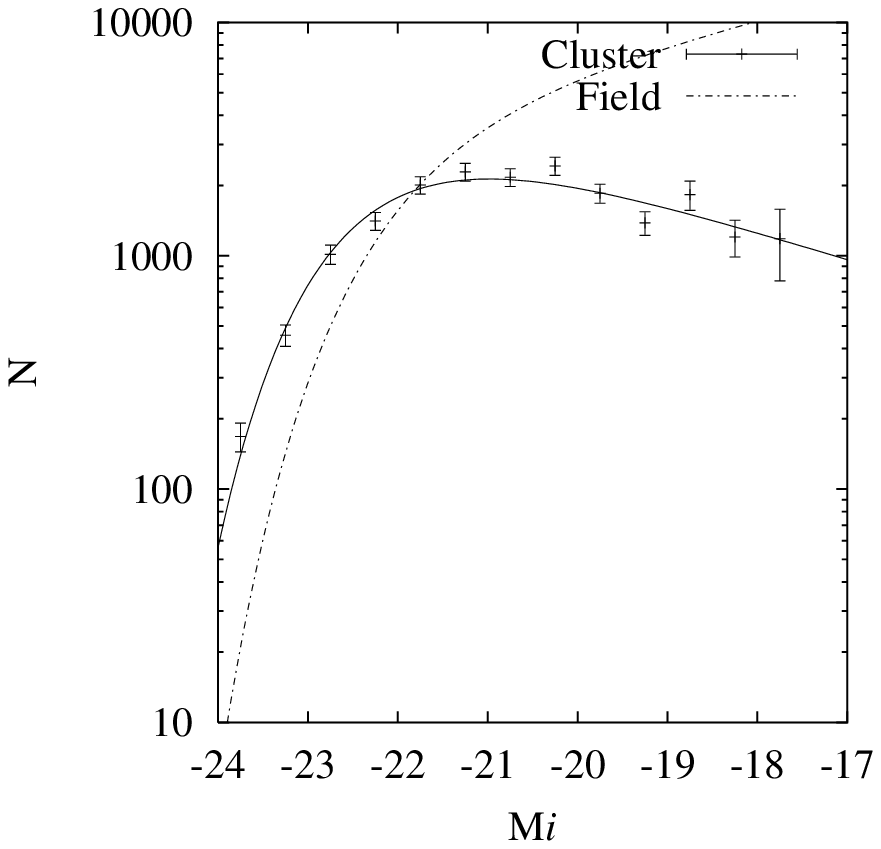}
\includegraphics[scale=0.5]{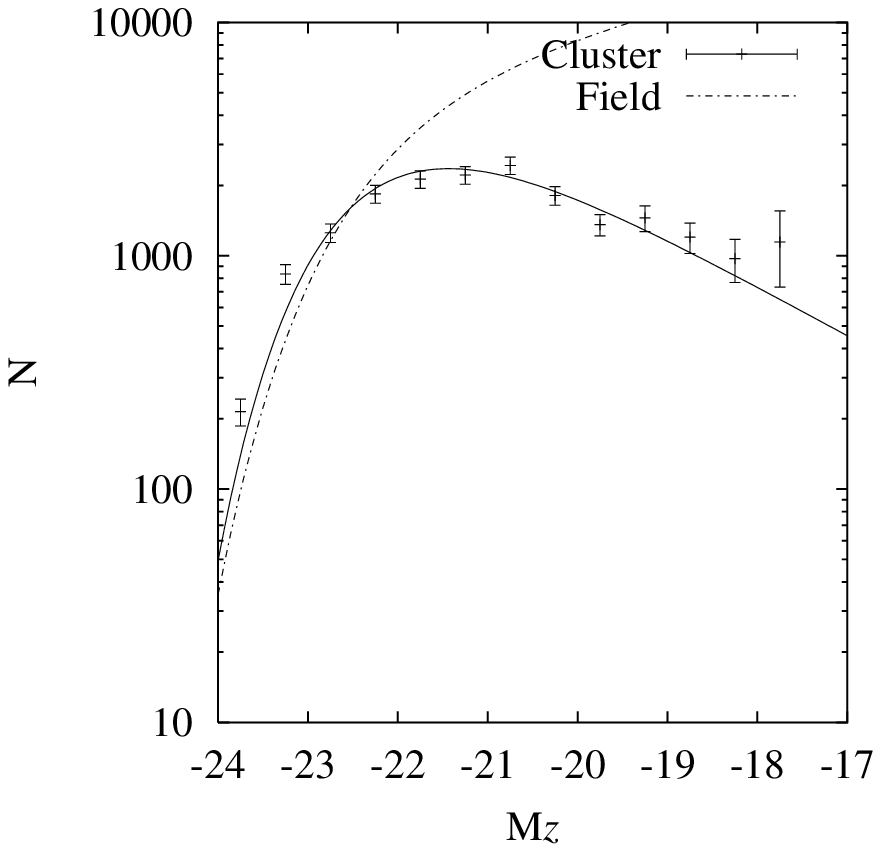}
\includegraphics[scale=0.5]{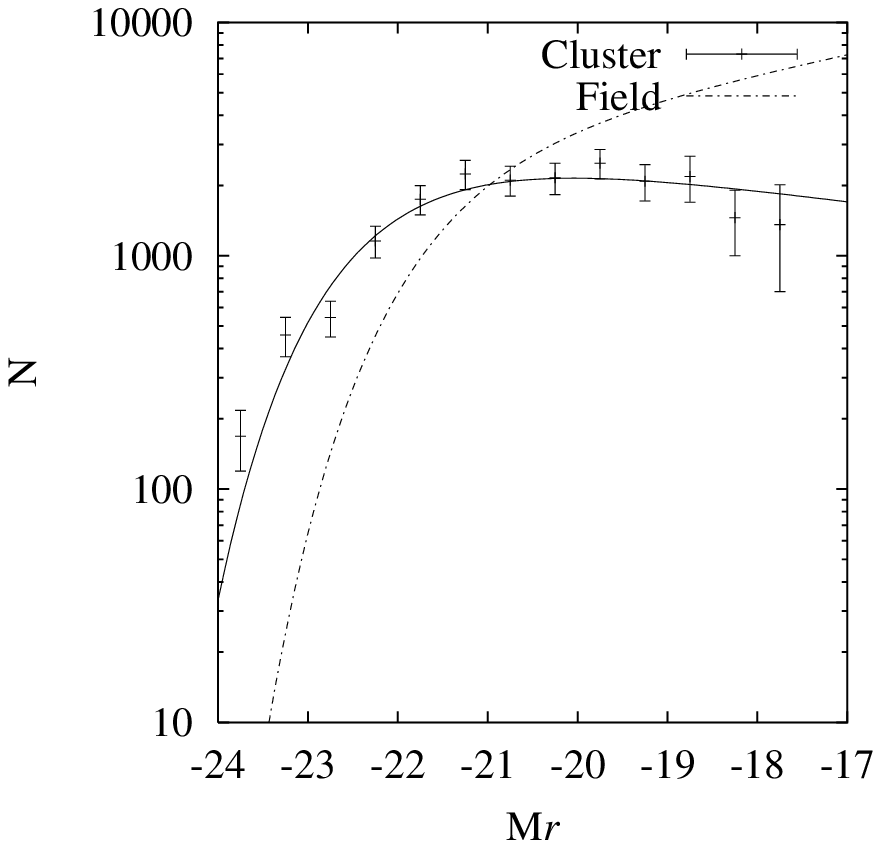}
\end{center}
\caption{
\label{fig:all.eps}
Composite LF of galaxy clusters from the SDSS CE galaxy cluster catalog
 in five SDSS bands. 
The solid line is the best-fit Schechter functions. The y-axis is arbitrary.
 The dotted line is the field LFs from Blanton et al. (2001) re-scaled to
 our cosmology. The normalization of field LFs was adjusted to match the
 cluster best-fit
 Schechter functions.
 The lower-right panel is for clusters with spectroscopic redshifts in
 $r$. The best-fit Schechter parameters are summarized in table \ref{tab:5color}.
}
\end{figure}

 In figure \ref{fig:all.eps}, we show the composite
 LF of the subset of CE clusters discussed above. We present one composite
 LF for each of the five SDSS passbands.  We also present in table
 \ref{tab:5color} the best-fit parameters from a joint fit of a Schechter
 function to these data. For a comparison, we also show the field values as
 derived by Blanton et al. (2001) (corrected for $h_0$ = 0.7). 
 In figure \ref{fig:all.eps}, field LFs normalized to cluster LFs are shown by dotted lines.
 As expected, the $M^*$ for our cluster LFs
 is significantly brighter (by 1 -- 1.5 magnitudes depending on the
 bands) than those seen for the field LFs in all five bands.
 Furthermore, the faint end slopes ($\alpha$) of the cluster composite
 LFs are much flatter that those seen for the field LFs. This is
 especially noticeable for the redder passbands ($i$ and $z$)
 while the slope of the cluster LF systematically flattens from the $u$
 passband to the $z$ passband. 

 These results are consistent with the hypothesis that the cluster LF has two
 distinct underlying populations i.e. the bright end of the LF is
 dominated by bright early types that follow a Gaussian-like luminosity
 distribution, while the faint-end of the cluster LF is a steep
 power-law-like function dominated by star-forming (bluer) galaxies.
 Binggeli et al. (1988) originally suggested this hypothesis, 
 while the recent work of Adami et al. (2000), Rakos et al. (2000) and
 Dressler et al. 1999 supported this idea. 
 Particularly, Boyce et al. (2001) showed LF of Abell 868 is made up
 of three different populations of galaxies; luminous red and two
 fainter blue populations. 
  The idea is illustrated by the fact that the cluster LFs in the redder passbands, which
 are presumably dominated by the old stellar populations of the early types,
 have much brighter $M^*$'s and significantly shallower slopes than those
 measured in the bluer passbands. Those results can also be interpreted
 as showing that bright elliptical galaxies are more populated 
 in dense regions, like inside of clusters. They are consistent with
 the morphology-density relation advocated by Dressler et al. (1980, 1997).

\subsection{The Composite Cluster LF as a Function of Morphology}\label{sec:morph}

 One of the key aspects of the SDSS photometric data is the opportunity to
 statistically study the distribution of galaxies as a function of their
 morphology. In this subsection, we discuss the composite cluster LF as a function
 of morphology using three complementary methods for determining the
 morphological type of each galaxy. These include:  i) the best-fit de
 Vaucouleur or exponential model profile; ii) the inverse of concentration
 index and iii) the $u-r$ color of the galaxies.
  We present all three methods, since at present it is unclear which
 method is the most successful at separating the different morphological galaxy
 types. Also, each method suffers from different levels of contamination, and
 the differences in the methods can be used to gauge the possible systematic
 uncertainties in the morphological classifications. We discuss the three
 methods used in detail below.

\begin{figure}[h]
\begin{center}
\includegraphics[scale=0.5]{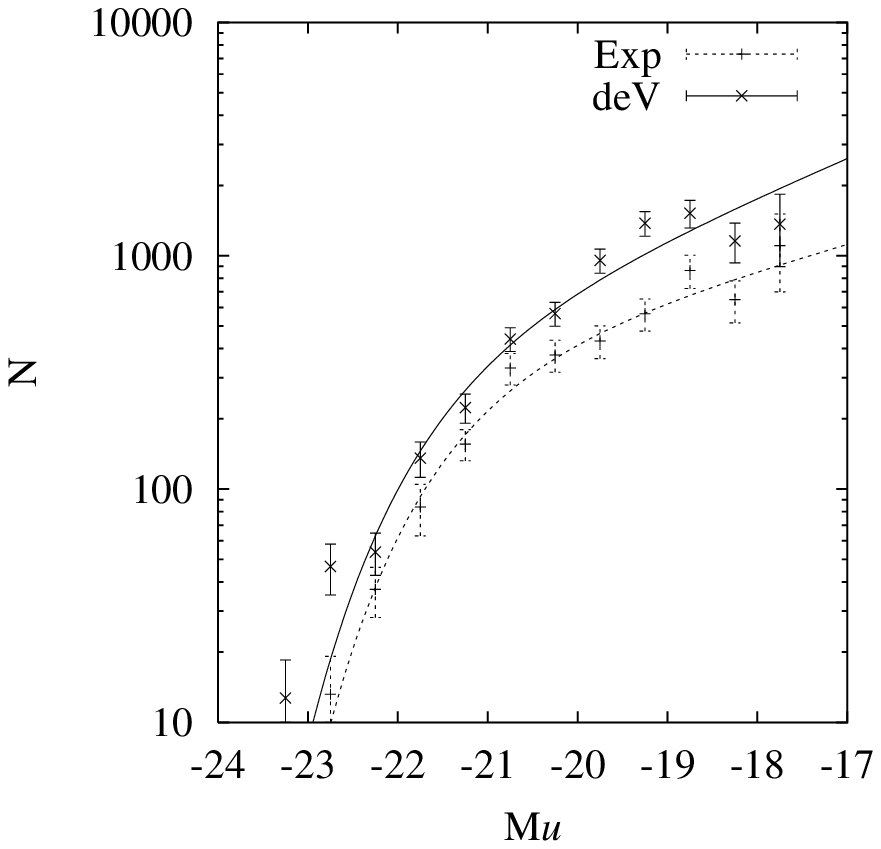}
\includegraphics[scale=0.5]{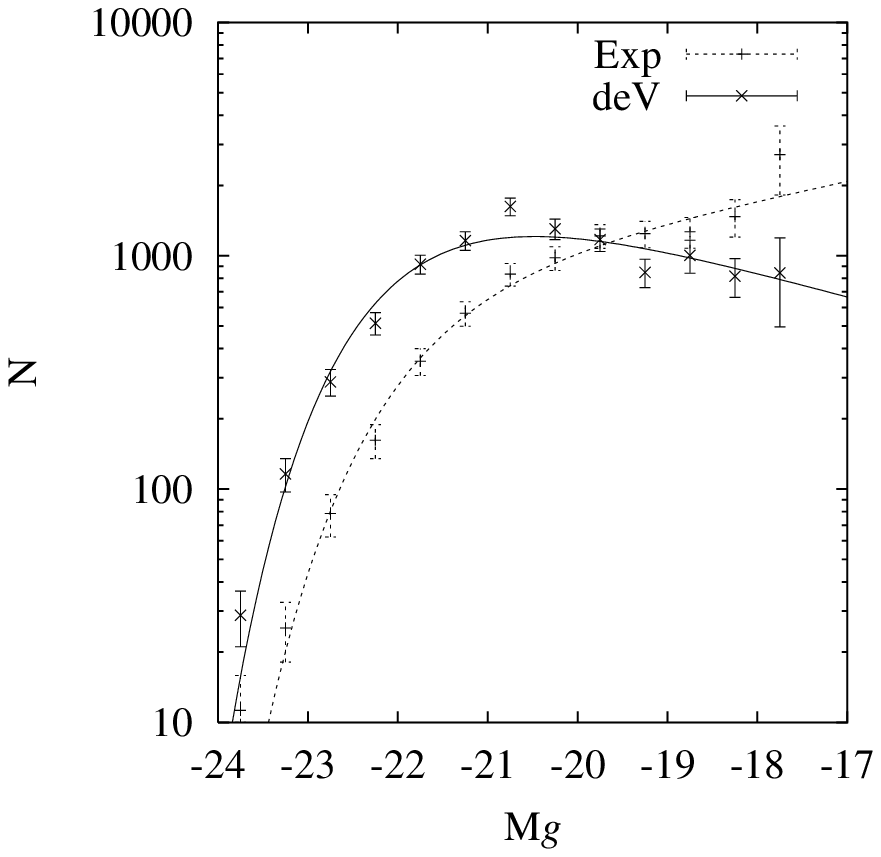}
\includegraphics[scale=0.5]{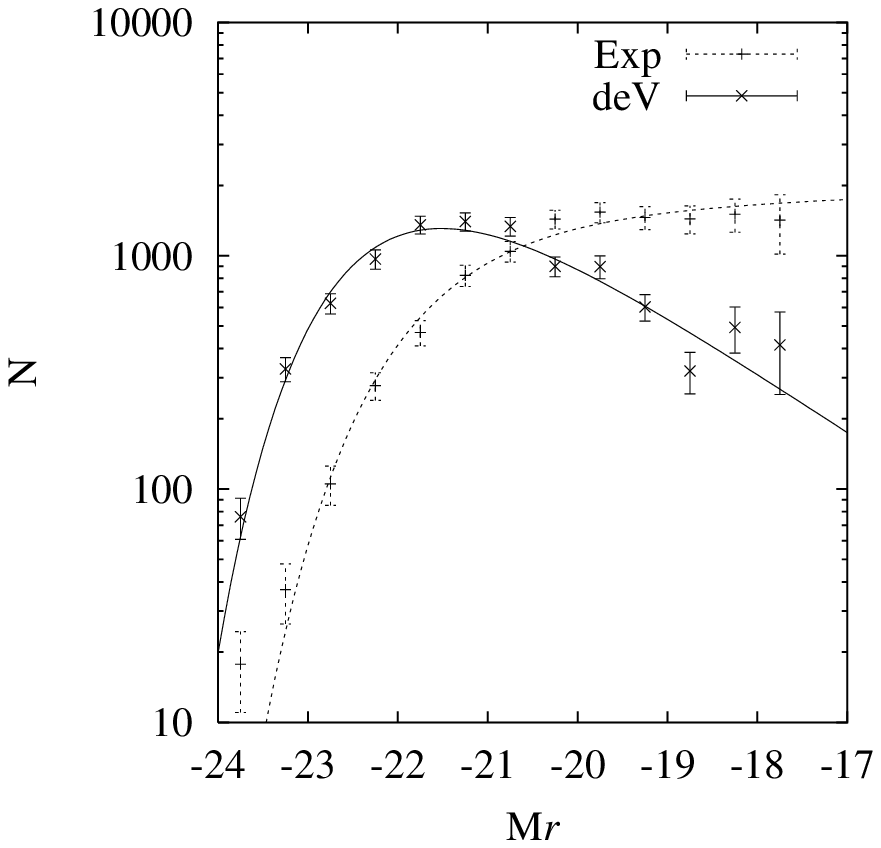}
\end{center}
\begin{center}
\includegraphics[scale=0.5]{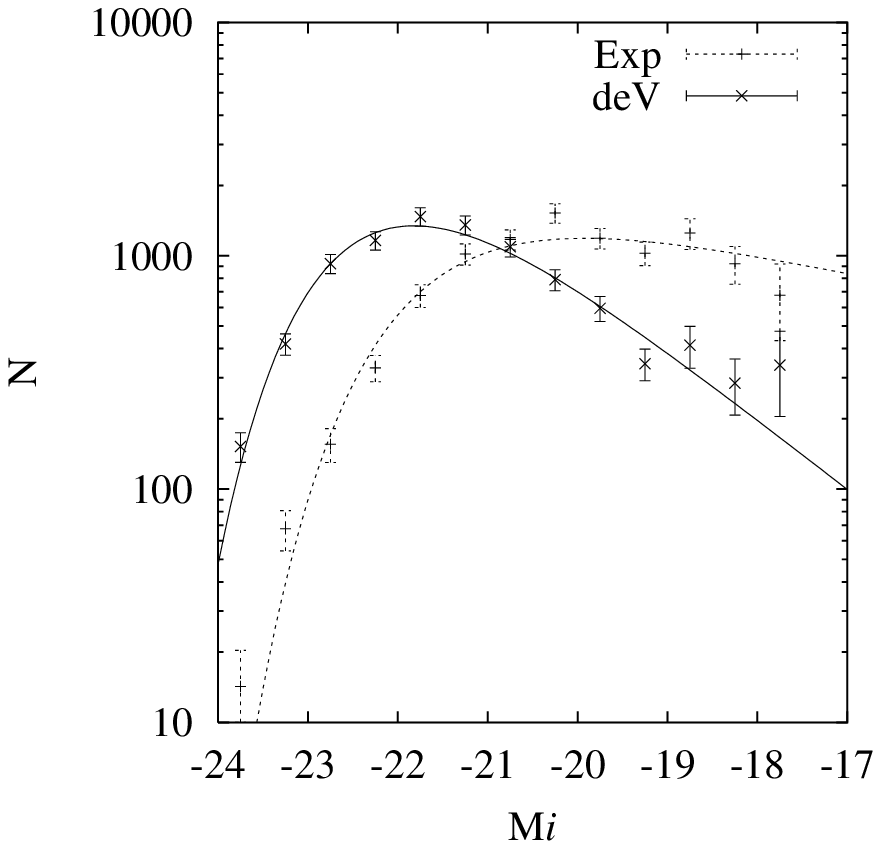}
\includegraphics[scale=0.5]{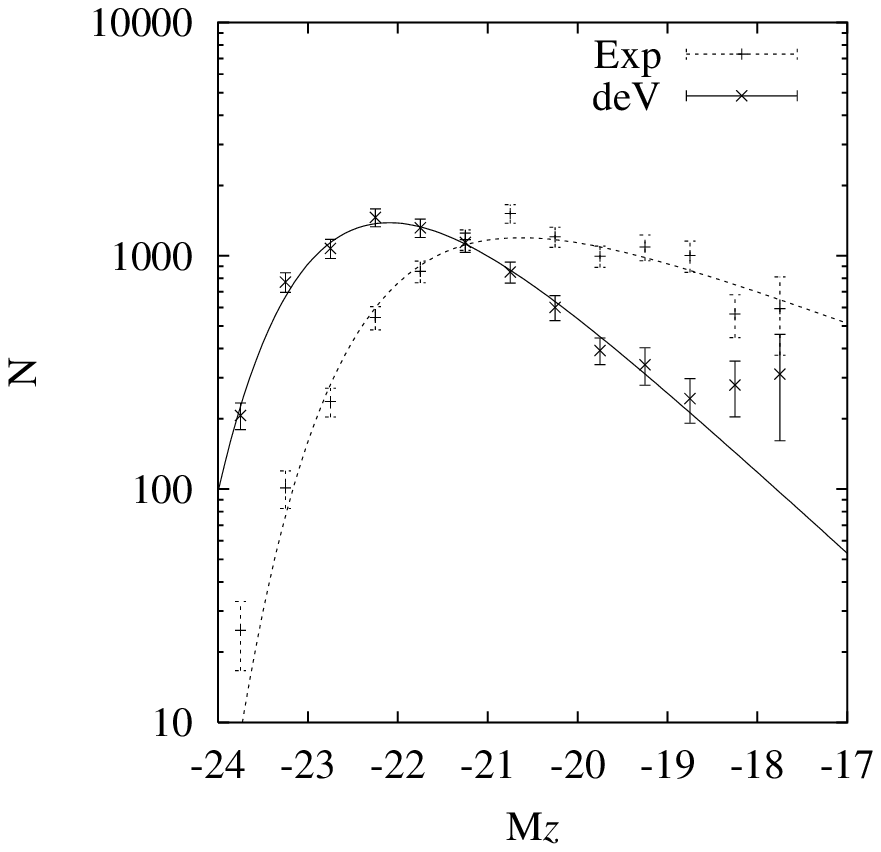}
\end{center}
\caption{
\label{fig:exp_dev.ps}
 Composite luminosity functions of de Vaucouleur galaxies and exponential
 galaxies. The galaxies are divided into two subsamples using profile
 fitting. The lines show the best-fit Schechter functions (solid for de
 Vaucouleur galaxies, dotted for exponential galaxies). The y-axis is arbitrary.
 de Vaucouleur galaxies always have a brighter $M^*$ and a flatter faint end
 tail. The best-fit Schechter parameters are summarized in
 table \ref{tab:dev_exp}.
}
\end{figure}

 The first method we consider here is using the de Vaucouleur and
 exponential model fits of the galaxy light profiles measured by SDSS photometric
 pipeline ($PHOTO$ R.H. Lupton et al., in preparation) to broadly
 separate galaxies into the late and early-type. If
 the likelihood of a de Vaucouleur model fit to the data is higher than the
 that of an exponential model fit, the galaxy is called a late-type, and vice versa. Galaxies that have the same likelihoods for both
 model fits are discarded. In figure \ref{fig:exp_dev.ps}, we present the
 composite cluster LF of late-type and early-type galaxies (as defined using
 the model fits above) for all five SDSS passbands. In table \ref{tab:dev_exp},
 we present the best-fit Schechter function parameters to these data, and show
 the fits in figure \ref{fig:exp_dev.ps}.

\begin{figure}[h]
\begin{center}
\includegraphics[scale=0.5]{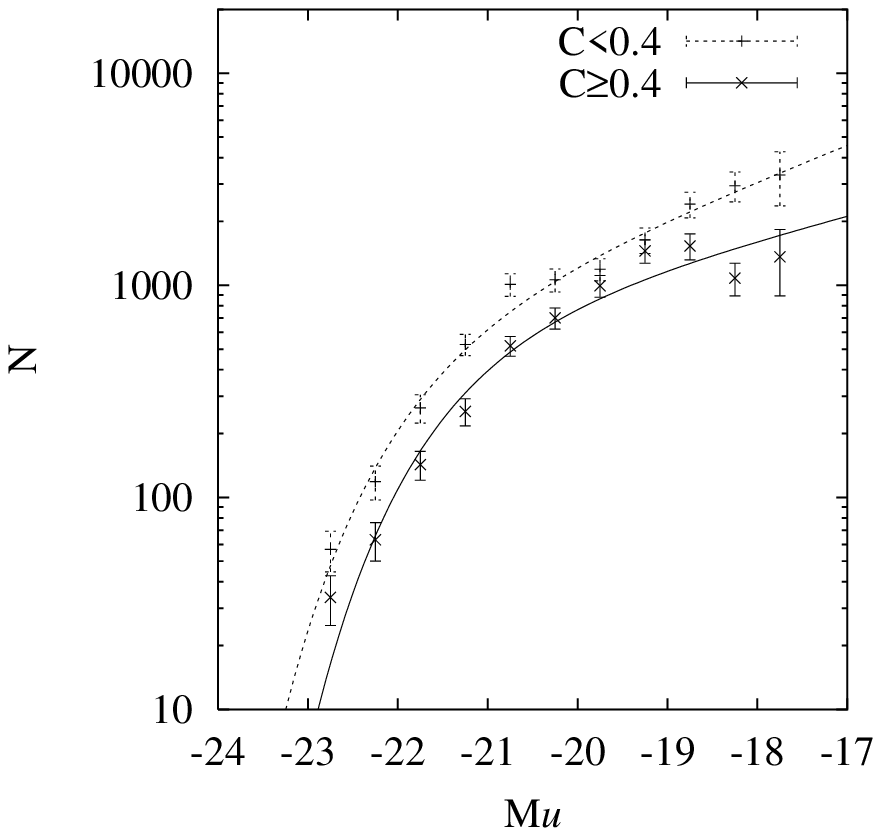}
\includegraphics[scale=0.5]{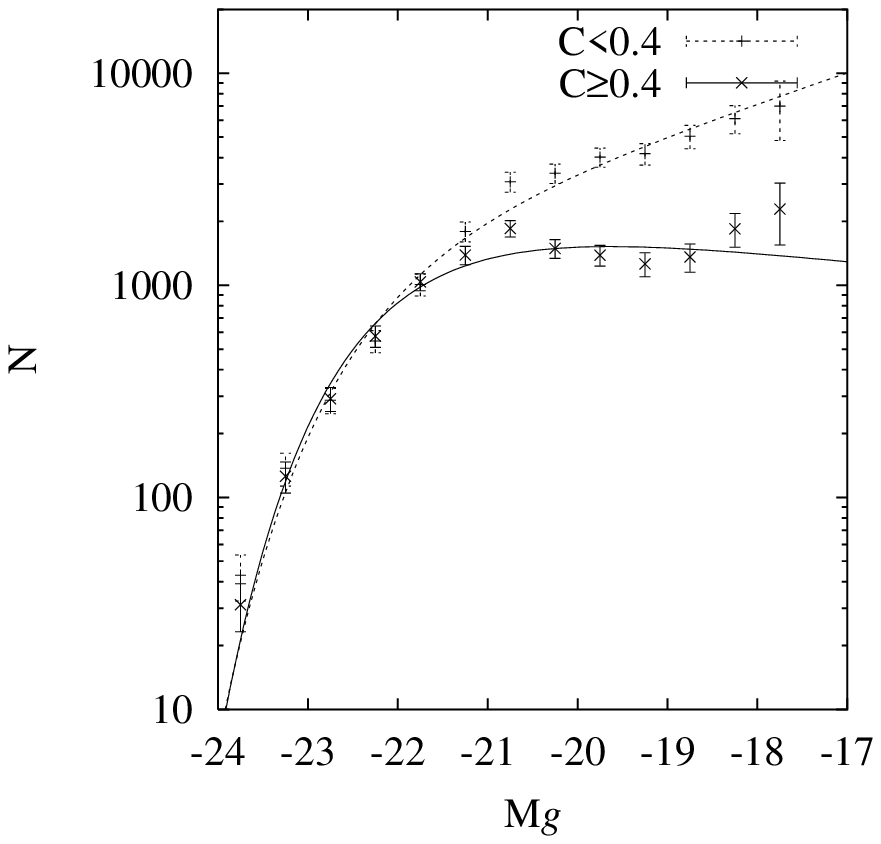}
\includegraphics[scale=0.5]{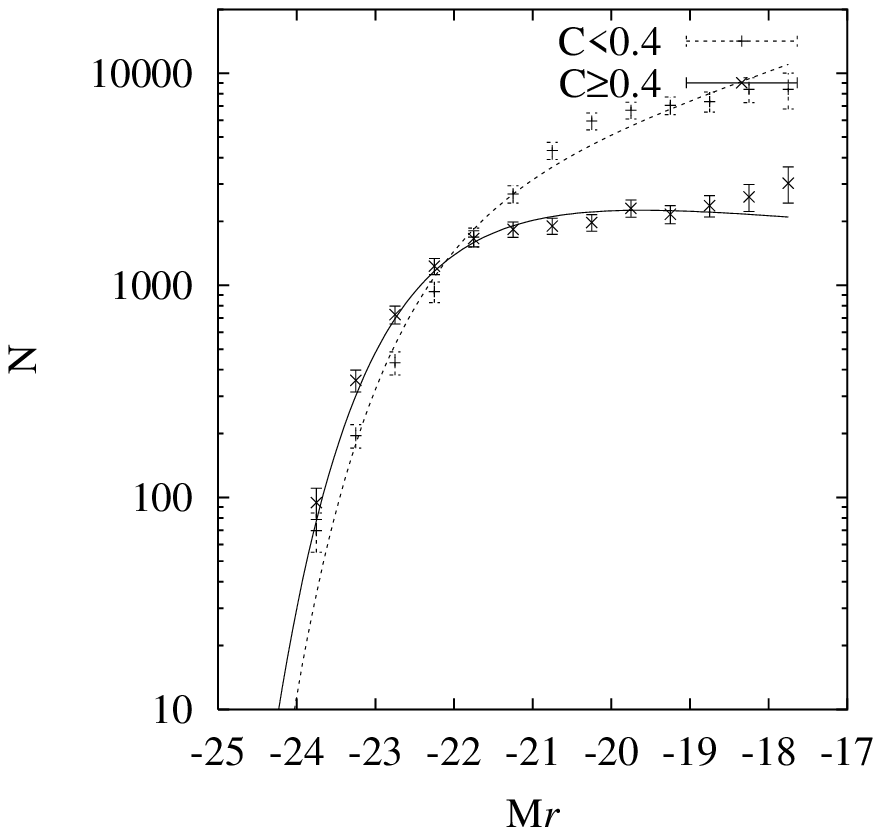}
\end{center}
\begin{center}
\includegraphics[scale=0.5]{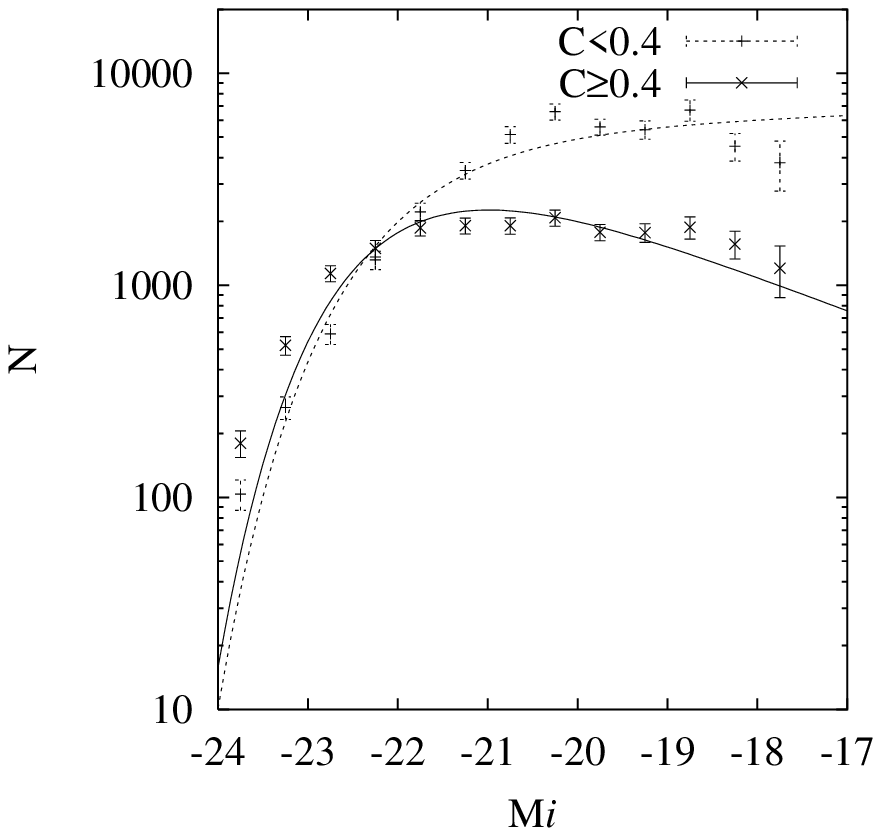}
\includegraphics[scale=0.5]{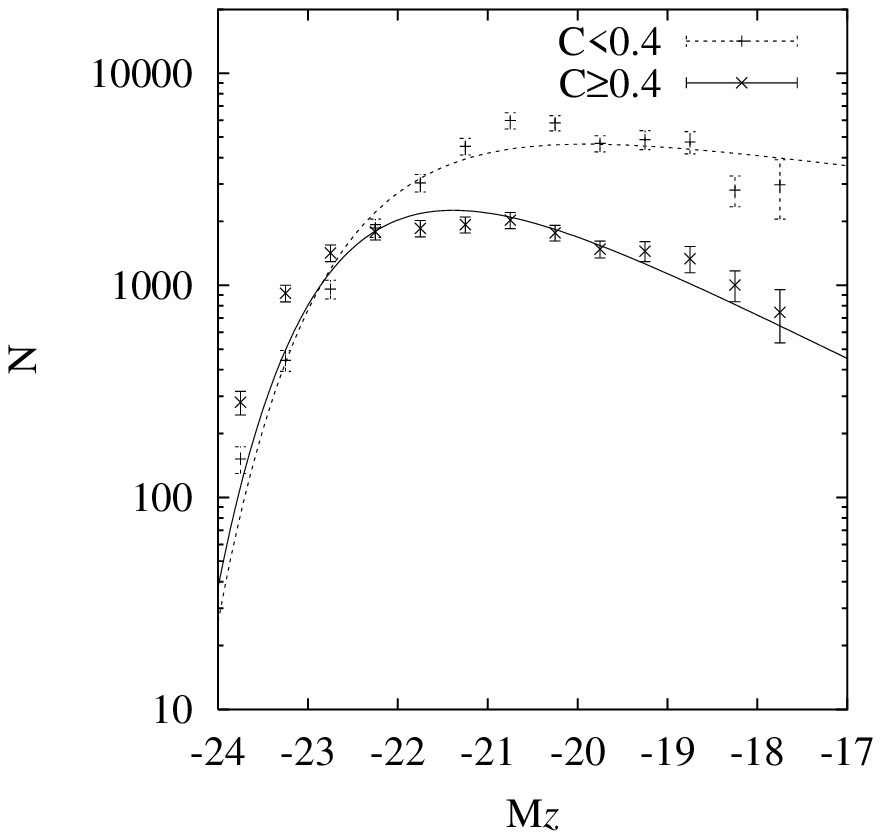}
\end{center}
\caption{
\label{fig:cin.ps}
Composite luminosity functions of high-concentration and low-concentration
 galaxies. The concentration index ($C$) used here is  the ratio of the 50\%
 Petrosian flux radius to the 90\% Petrosian flux radius. 
 In this figure, early-type galaxies have $C<$0.4, and
 late-type galaxies have $C\geq$0.4. Early-type galaxies
 have flatter faint end tails in all five bands. 
 Lines are the best-fit Schechter functions (solid for $C<$0.4, dotted
 for $C\geq$0.4). The y-axis is
 arbitrary. 
 The best-fit Schechter parameters are summarized in
 table \ref{tab:5color_cin}.
}
\end{figure}

  The second method uses the inverse of the concentration index, which is defined
 as $C$ = $r_{50}/r_{90}$, where $r_{50}$ is the radius that contains
 50\% of the Petrosian flux and $r_{90}$ is the radius that contains
 90\% of the Petrosian flux (see R.H. Lupton et al., in preparation). Both
 of these parameters are measured by the SDSS PHOTO analysis pipeline for each
 galaxy. The concentration parameter used here ($C$) is just the inverse of the
 commonly used concentration parameter, and thus early-type galaxies have a
 lower $C$ parameter than late-type galaxies.  The correlation of
 $C$ with visually-classified morphologies has been studied in detail by
 Shimasaku et al. (2001) and Strateva et al. (2001). They found that galaxies
 with $C<$0.4 are regarded as early-type galaxies, while galaxies with
 $C\geq$0.4 are regarded as late-type galaxies.  Therefore, in figure
 \ref{fig:cin.ps}, we show the composite cluster LF of late-type and
 early-type galaxies as defined using this second method for all five SDSS
 passbands. In table \ref{tab:5color_cin}, we present the best-fit Schechter
 function parameters to these data.

\begin{figure}[h]
\begin{center}
\includegraphics[scale=0.5]{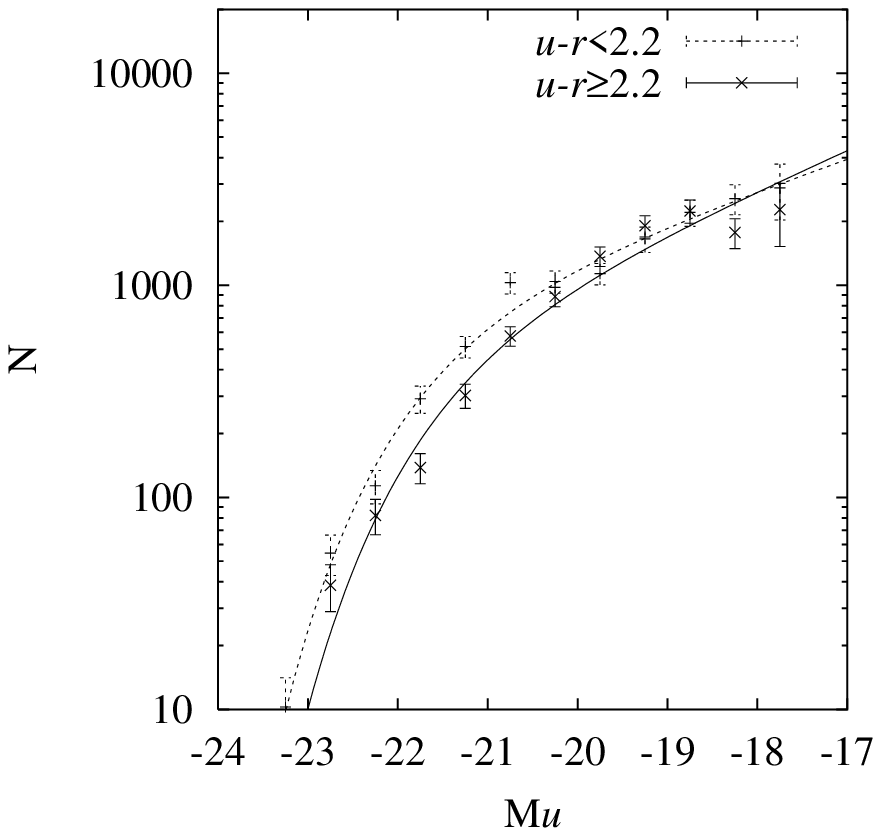}
\includegraphics[scale=0.5]{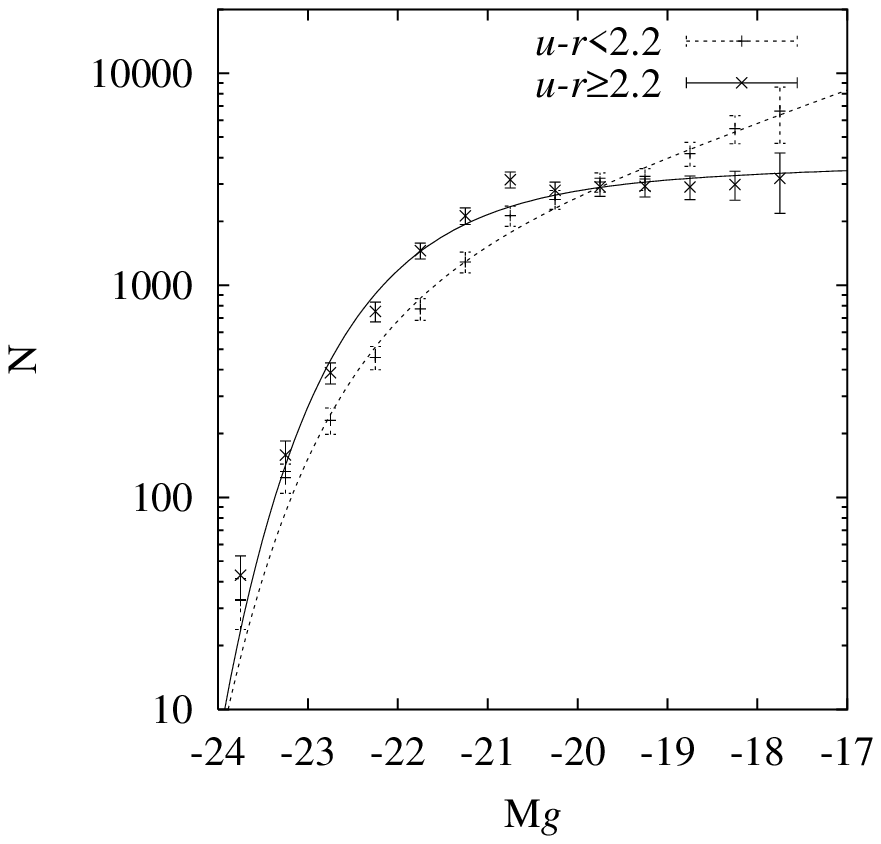}
\includegraphics[scale=0.5]{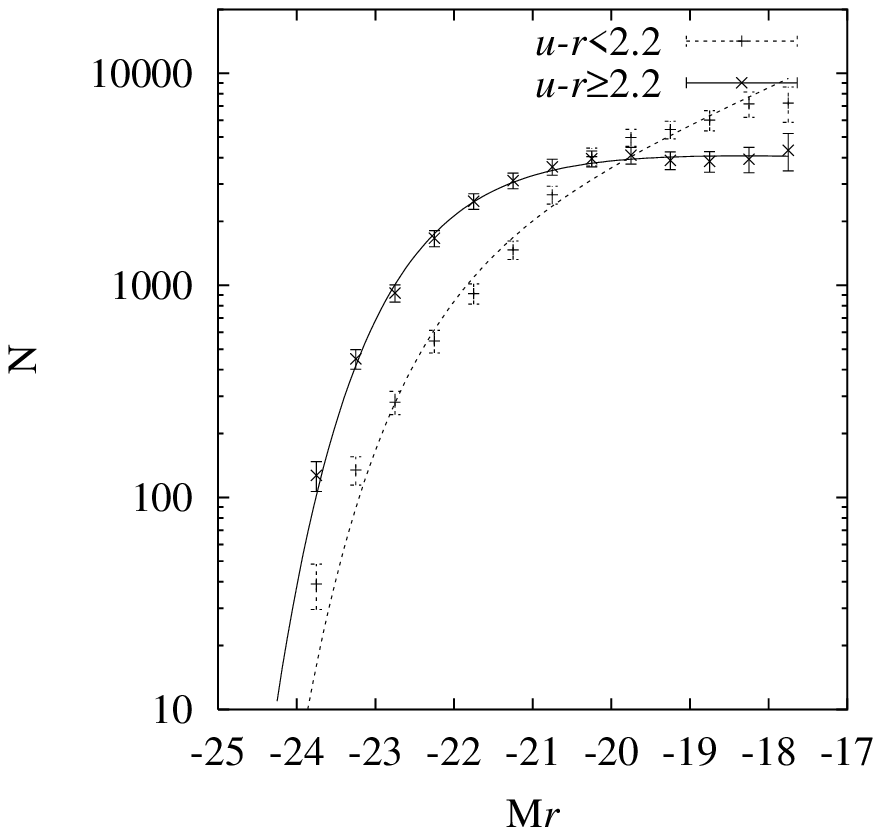}
\end{center}
\begin{center}
\includegraphics[scale=0.5]{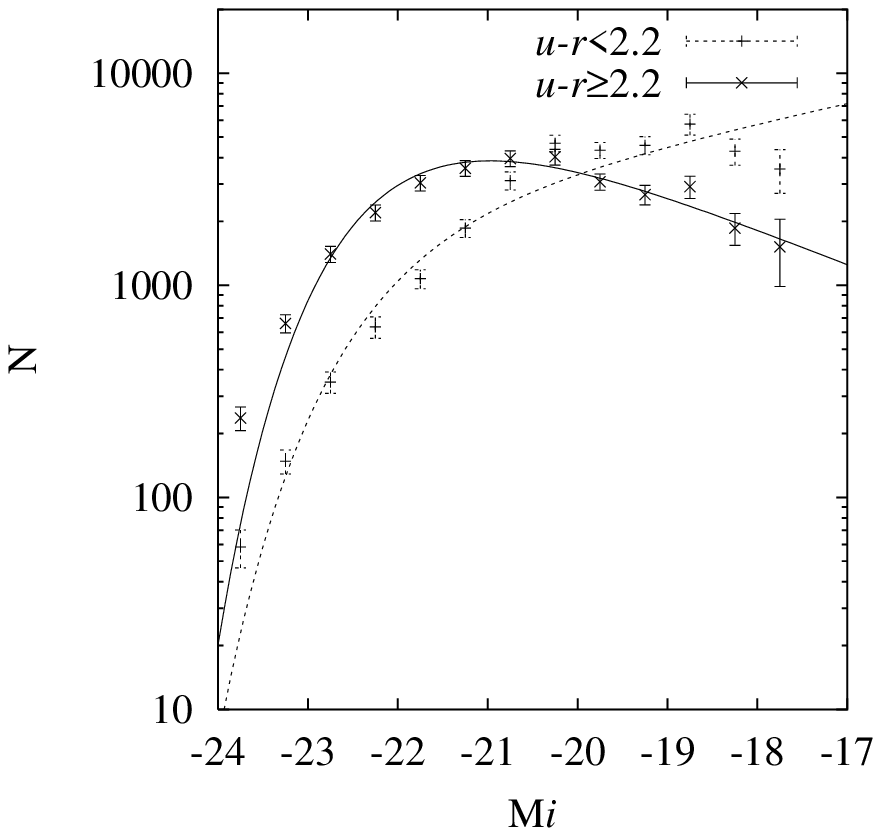}
\includegraphics[scale=0.5]{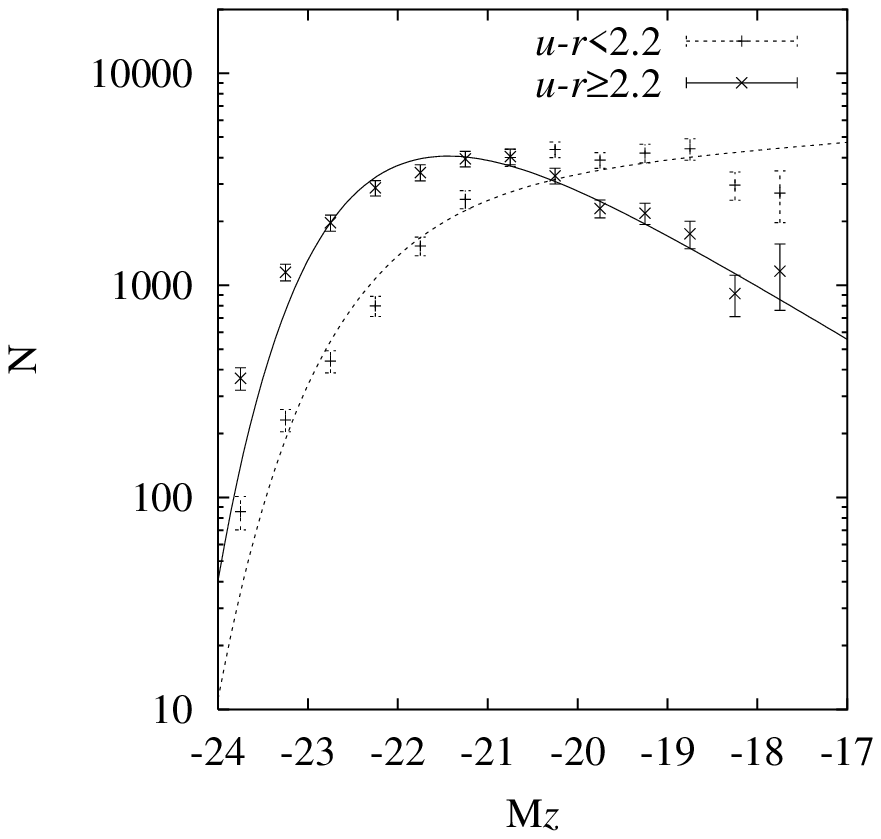}
\end{center}
\caption{
\label{fig:ur.ps}
Composite luminosity functions of $u-r<$ 2.2 (late-type) and
 $u-r\geq$ 2.2 (early-type)
 galaxies. Early-type galaxies
 have flatter faint end tails in all five bands. 
 The lines are the best-fit Schechter functions (solid for $u-r<$ 2.2,
 dotted for $u-r\geq$ 2.2). The y-axis is
 arbitrary.  The best-fit Schechter parameters are summarized in table \ref{tab:5color_ur}.
}
\end{figure}

 The third method used herein for morphological classification was to use the
 observed $u-r$ color of the galaxy which has been proposed by Strateva et
 al. (2001). Using the fact that k-correction for $u-r$ is almost
 constant until $z$ = 0.4, they showed that galaxies shows a clear bimodal distribution in
 their $u-r$ color and $u-r$ = 2.2 serves as a good classifier of morphology
 until $z\sim$ 0.4 by correlating $u-r$ classification with visual
 classifications. Therefore, we have classified galaxies with $u-r<$ 2.2 as
 early-type and galaxies with $u-r\geq$ 2.2 as late-type.  Figure
 \ref{fig:ur.ps} shows the composite cluster LF for both types of galaxies
 along with their best-fit Schechter functions (in all five passbands). The
 best-fit Schechter parameters are summarized in table \ref{tab:5color_ur}.

 As expected, there are  noticeable differences in these three morphological
 classifications, as portrayed by the differences in their composite LFs (see
 figures. \ref{fig:exp_dev.ps}, \ref{fig:cin.ps}, and \ref{fig:ur.ps}).  However,
 it is worth stressing here the similarities between the methods. For example,
 the faint end slope of the LF is always shallower for early-type galaxies
 than late-type, regardless of the passband and methodology. Also, the faint end
 slope for early-type galaxies decreases steadily toward the redder
 passbands, while the faint-end slope for the late-type galaxies is nearly
 always above $-1$ and consistent (or steeper) than the field LF in most
 passbands. These observations are again qualitatively in agreement with the
 hypothesis that the bright end of the cluster LF is dominated by bright, old
 early-types, while the faint-end of the
 cluster LF represents late-type galaxies maybe in greater numbers
 than the average field. This model is in agreement with hierarchical models of
 structure formation and the model for the tidal disruption of dwarf galaxies
 by the dominant early types.

\section{Discussion}
\label{discussion}

 In this section, we discuss various tests which we have performed on our
 measurement and results.

\subsection{Monte-Carlo Simulations}\label{sec:monte}

 To test the robustness of our methods, we performed Monte-Carlo
 simulations which involved adding artificial clusters to the SDSS data and
 computing their composite LF using the same algorithms and software as used on
 the real data.  Our model for the artificial clusters was constructed using
 the SDSS data on Abell 1577 (at $z\sim0.14$, richness$\sim1$). We used
 the method described in Goto et al. (2002) to make artificial clusters. 
 The radial profile for the artificial clusters was taken to be a King profile (Ichikawa
1986) with a concentration index of 1.5 and a cut-off radius of 1.4 Mpc, which is the size of Abell 1577 (Struble, Rood 1987).  The
 color-magnitude distributions for the artificial clusters were set to be the
 observed, field-corrected, color-magnitude distributions of Abell 1577
 binned into 0.2 magnitude bins in both color and magnitude.
 From this model, we then constructed artificial clusters as a function of
 the redshift and overall richness. For redshift, we created clusters at
 $z$ = 0.2, 0.3, 0.4 and 0.5, ensuring that we properly accounted for the cosmological
 effects; i.e. the clusters became smaller, redder and dimmer with
 redshift.  We used the k-corrections for an early type spectrum.  For
 richness, we change the number of galaxies within each cluster
 randomly between 10 and 50. For each redshift, we
 created 100 clusters (400 clusters in total). The galaxies within these
 artificial cluster were distributed randomly in accordance with the radial and
 color-magnitude distributions discussed above. We made no attempt to
 simulate the density-morphology relation nor the luminosity segregation in
 clusters.

\begin{figure}[h]
\begin{center}
\includegraphics[scale=0.7]{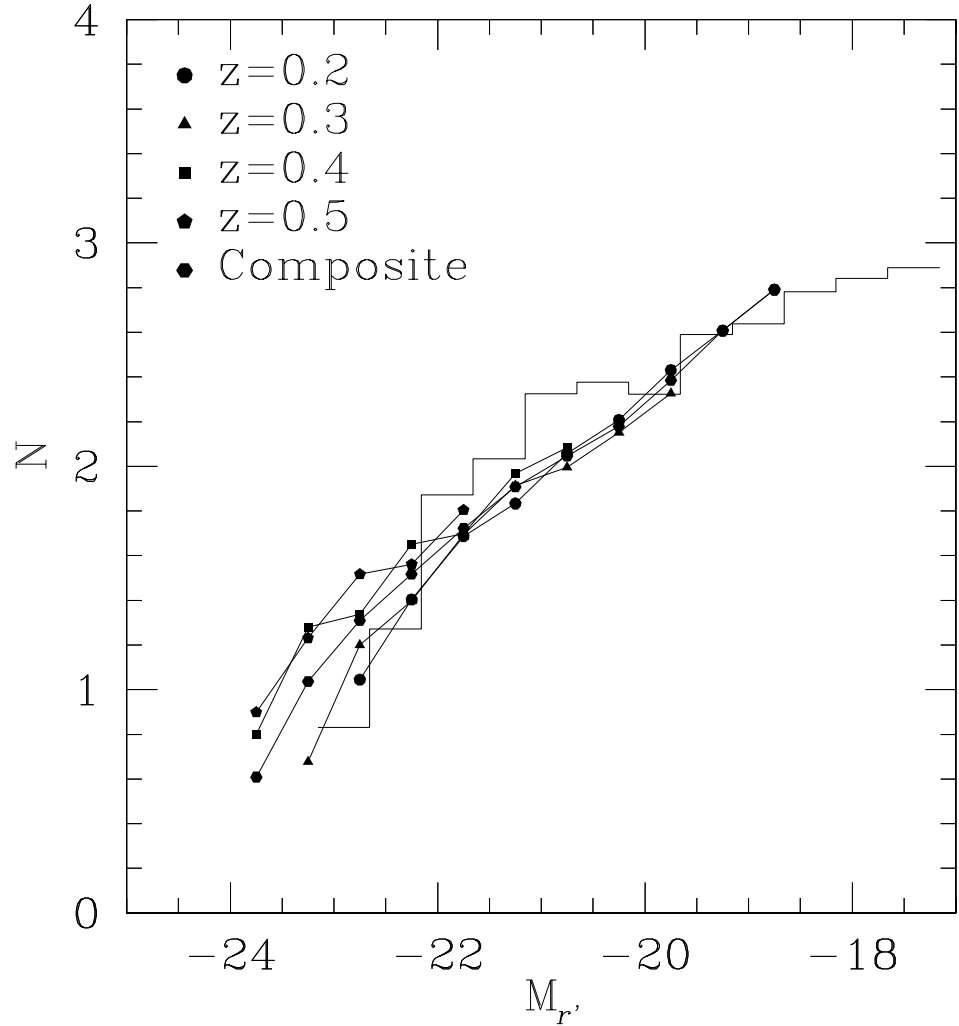}
\end{center}
\caption{
\label{fig:tfake_1577_ab.ps}
 Results of a Monte-Carlo simulation to test the robustness of the
 weighting scheme. 
 The histogram shows the luminosity function of model cluster A 1577.
 The Circles, triangles, squares and pentagons represent the
 composite luminosity function at each redshift ($z$ = 0.2, 0.3. 0.4 and
 0.5, respectively) constructed with 100 fake clusters at each
 redshift. The hexagonals show the composite luminosity function from all 400 fake
 clusters distributed on the real SDSS data. 
}
\end{figure}

 The artificial clusters were randomly distributed within the real SDSS imaging
 data and we constructed a composite LF for these clusters using exactly the same
 software as for the real clusters. Since the artificial clusters were all made
 from the same luminosity distribution, the composite LF should therefore look
 very similar in shape as the original input LF. Figure
 \ref{fig:tfake_1577_ab.ps} shows the  result of our Monte-Carlo
 simulations. The histogram shows the original absolute magnitude
 distribution of Abell 1577 after a field correction, while the symbols show
 the composite luminosity functions which we constructed as a function
 of the input redshift.

\subsection{Check of Photometric Redshifts}

 One of the most innovative parts of this analysis is the use of photometric
 redshifts to determine the composite luminosity function of clusters. As
 demonstrated in Goto et al. (2002), the accuracy of photometric redshift is
 excellent ($\delta z$ = $\pm$0.015 for $z<$0.3) and this method will certainly
 be used in the near future as the number of clusters with photometric
 redshifts will increase rapidly, far quicker than the number of clusters with
 spectroscopically confirmed redshifts. 

 To justify our use of photometric redshifts, since all previous composite
 cluster LF's used spectroscopic redshifts, we constructed a
 composite LF using only the clusters with spectroscopically confirmed
 redshifts. We derived our spectroscopic redshift for CE clusters by matching
 the SDSS spectroscopic galaxy data with our CE clusters. This was achieved by
 searching the SDSS spectroscopic galaxy sample for any galaxies within the CE
 cluster radius and within $\delta z$ = $\pm$0.01 of the photometric redshift of the
 cluster. 
 The radius used here was from Goto et al. (2002). If multiple
 galaxies satisfied this criteria, the closest spectroscopic redshift to the photometric
 redshift was adopted. The number of clusters with spectroscopic redshifts
 was 75 out of 204 at the date of this writing.

 The results of this test are shown in figure \ref{fig:all.eps}
 (in the bottom right-hand panel).  Also the parameters for the best-fit
 Schechter functions are given in table \ref{tab:5color}.
 and referred to as
 $r^*$(spec). We only performed this test for the $r$
 passband. The slope and characteristic magnitude of the best-fit Schechter
 function for the spectroscopically determined LF is in good agreement with
 that derived using photometric redshifts. As can be seen in table \ref{tab:5color},
 both $M_r^*$ and the slope agree within the error. This test shows that
 we can truly construct composite LFs using photometric redshift of clusters.

\subsection{Test of Cluster Centroids}
 
 One key aspect of measuring the composite cluster LF is the choice of
 the cluster centroid.
   To test the effect of different cluster centroids on the composite LF, we constructed a
 composite cluster LF using the position of Brightest Cluster Galaxies (BCGs) as a
 centroid instead of the peak in the enhanced density map, as discussed in Goto
 et al. (2002). The BCGs have been determined to be the brightest galaxy
 among galaxies fainter than $-$24th magnitude within 0.75 Mpc. 
 Galaxies brighter than $-$24th magnitude are regarded as being foreground galaxies.
 The mean offset between the BCG position and the centroid previously used is 1.02 arcmin. 
  Table \ref{tab:bcg_center} lists the parameters of the best-fit Schechter functions
 to the five SDSS passbands, which should be compared to the values obtained
 using the optical centroid given in table \ref{tab:5color}.
 In all five bands, the characteristic magnitudes and slopes agree very
 well within the error. 
 This test shows that our composite LFs are not dependent on a center determination.

\subsection{Test of Background Subtraction}

 Since we constructed composite LFs from 2-dimensional, projected sky image, 
 subtraction of fore/background galaxies played an important role in this work.
 We test here the effect of making a global background subtraction for all
 clusters instead of the local background subtraction discussed above. 
 We use the number-magnitude relation of all the galaxies in the entire
 150 deg$^2$ region as the global background.  Table \ref{tab:global} gives
 the best-fit Schechter parameters of composite LFs constructed using
 global background subtraction.
 Compared with table \ref{tab:5color},
 again, every Schechter parameter agrees very well within 1 $\sigma$.
 Although we use annuli around clusters to subtract the background to
 avoid the large-scale structure disturbing the measurement of composite
 LFs, this test shows that our composite LFs are not dependent on
 background subtraction. Valotto et al. (2001) showed that
 a statistical background subtraction can not re-produce composite LFs
 using a mock galaxy catalog constructed from a large $N$-body
 simulation. Our result, however, combined with the fact that we derived the
 same LF as input through a Monte-Carlos simulation (in subsection
 \ref{sec:monte}), supports that our composite LFs are not subject to
 background subtraction.

\subsection{Test of Cluster Richness}

 Another aspect we were concerned about was our choice of cluster richness criteria.
 To test this, we constructed composite LFs of different subsample with
 $N_{-18}>$20 and    $N_{-18}>$40, given in table \ref{tab:richness}. 
 $N_{-18}$ here is defined as the number of galaxies brighter than $-$18th
 magnitude after subtracting the background in the way we construct
 composite LFs.  $N_{-18}>$20 was used to
 construct composite LFs, as mentioned in section \ref{method}.  In table
 \ref{tab:richness}, even though $M^*$ is slightly brighter and
 the slope is slightly steeper for the richer sample, they agree within 1 $\sigma$. 
 The steepening of the slopes can be interpreted as a bias in selecting
 richer systems using $N_{-18}$,  i.e. Clusters with steeper
 tails tend to have a larger value of $N_{-18}$.
 This, however, confirms  that our composite LFs are not dependent on
 the  richness criteria, which we chose.

\subsection{Comparison with Other LFs}

 As the final test of our composite cluster LF, we compare here our composite
 LFs with previous works. First, we must be careful to match the different
 cosmologies used by various authors as well as the different photometric
 passbands. To facilitate such a comparison, therefore, we present in table \ref{tab:previous} the best-fit Schechter function parameters for our
 composite LF, but calculated for each author's cosmology and passband using the
 color corrections of Fukugita et al. (1995) and Lumsden et al. (1992).

 In the case of the three $b_j$ photographic surveys of Colless (1989),
  Valotto et al. (1997),  and Lumsden et al. (1997), we find a significantly brighter $M^*$ than these
 studies as well as a much shallower slope.
 We also tried to fit a Schechter function using their $\alpha$ value
  for the slope, but $M^*$'s become even brighter. The fits are not 
 good when fixed $\alpha$'s are used. 

 Lugger (1989) found   $M_R$=$-$22.81$\pm$0.13  and
 $\alpha$=$-$1.21$\pm$0.09 by re-analyzing nine clusters presented in
 Lugger (1986). The slope is steeper  and $M^*$ is
 slightly brighter than our results. When we fix the slope with her
 value at $\alpha$=$-$1.21, the two LFs agree well.

 Garilli et al. (1999) studied 65 Abell and X-ray selected samples of galaxies
 in the magnitude range of -23.0$<M_r<$-17.5  and found that
 $M_r^*$=$-$22.16$\pm$0.15 and $\alpha$=$-$0.95$\pm$0.07 (in isophotal
 magnitudes).
  This slope is steeper
 than ours. A possible difference with ours is that they used the color condition to select
 cluster galaxies. $M^*$ is in agreement with our results within the
 error. We also tried to fit a Schechter function with a fixed value of
 $\alpha$ = 0.84. $M^*$ became brighter by 0.18 mag although the fit was poor.  
 
 Paolillo et al. (2001) studied composite LF of 39 Abell clusters using the
 digitized POSS-II plates. They obtained $M^*$=$-$22.17$\pm$0.16 in $r$.
 The slope is $\alpha$=$-$1.11$_{-0.09}^{+0.07}$.
  Although the slope differs significantly,
 $M^*$ agrees well compared with our composite LF.

 Yagi et al. (2002a,b) observed 10 nearby clusters with their Mosaic CCD
 camera to derive composite LF. Their best-fit Schechter parameters are
 $M^*$=$-$21.1$\pm$0.2 and  $\alpha$=$-$1.49$\pm$0.05 in $R$. 
 They also studied type-specific LF using exponential and $r^{1/4}$
 profile fits to classify the galaxy types. They derived $M^*$=$-$21.1
 and  $\alpha$=$-$1.49 for exponential galaxies and $M^*$=$-$21.2
 and  $\alpha$=$-$1.08 for $r^{1/4}$ galaxies. Considering that they
 derived composite LFs using the data taken with different instruments
 analyzed in a different way, it is reassuring that they
 reached the same conclusion as our results discussed in subsection
 \ref{sec:morph}. i.e. exponential galaxies have the steeper faint end
 tail than $r^{1/4}$ galaxies, while their $M^*$ are almost the same.

 Concerning the disagreement of our LFs with previous studies, various differences
 in measuring composite LFs may be the reason.
  The possible sources of differences are different ways of weighting, different ways of
 background subtraction, and different depths of the luminosity function. 
 The sample clusters, themselves, should have, to some extent, different
 richness distributions.
 For $M^*$, although we tried to transform our magnitude into their
 magnitude, the color conversion between SDSS bands and others might not be
 accurate enough.  Thus, the difference with the previous studies is not
 necessarily a mistake in the analysis, but rather it represents a different way of
 analysis. Throughout our analysis discussed in section \ref{method} we carefully
 used exactly the same way to construct the composite LFs. We thus keep our
 composite LFs internally consistent.

\section{Conclusions}\label{conclusion}

 We studied the composite LF of 204 the SDSS CE galaxy clusters.  The over-all
 composite LF is compared with other composite LFs.  Comparing it to the field
 luminosity function, a tendency of a brighter $M^*$ and a flatter slope is
 seen. This is consistent with our understanding that cluster regions are
dominated by brighter galaxies than field galaxies.
 We divided the composite LF by galaxy morphology in three ways. In all three
cases, we found that early-type galaxies have flatter slopes than
late-type galaxies. 
 These observations are in agreement with the
 hypothesis that the bright end of the cluster LF is dominated by bright, old
 early-types, while the faint-end of the
 cluster LF represents late-type galaxies.
 This is also consistent with the morphology--density relation
originally advocated by Dressler et al. (1980, 1997). 
 We also studied these composite
LFs in five SDSS color bands. The slopes become flatter and flatter
toward the redder color bands. This again suggests that cluster regions are
dominated by elliptical galaxies with old stellar population.
These composite LFs provide a good low redshift benchmark
to study higher-redshift clusters in the future.
 Since the data used in this work came from 2\% of the SDSS data, further studies with
large SDSS data will increase the statistical significance on these topics as
the SDSS proceeds.

\bigskip

 We would like to thank the referee, Steven Phillipps, for a detailed
 revision and useful recommendations provided for this work. 
 We are grateful to Michael Crouch and Christopher J. Miller for valuable
 comments, which contributed to improve the paper.  
  T.G. wishes to thank Robert C. Nichol for hospitality during the stay at Carnegie Mellon University.
 T. G. acknowledges financial support from the Japan Society for the
 Promotion of Science (JSPS) through JSPS Research Fellowships for Young
 Scientists. 

 The Sloan Digital Sky Survey (SDSS) is a joint project of The
University of Chicago, Fermilab, the Institute for Advanced Study, the
Japan Participation Group, the Johns Hopkins University, the
Max-Planck-Institute for Astronomy, New Mexico State University,
Princeton University, the United States Naval Observatory, and the
University of Washington. Apache Point Observatory, site of the SDSS
telescopes, is operated by the Astrophysical Research Consortium
(ARC).  Funding for the project has been provided by the Alfred
 P.~Sloan Foundation, the SDSS member institutions, the National
Aeronautics and Space Administration, the National Science Foundation,
the U.~S.~Department of Energy, Japanese Monbukagakusho, and the Max Planck
Society. The SDSS Web site is $\langle$http://www.sdss.org/$\rangle$.


\newpage

\clearpage

\clearpage

\clearpage

\clearpage

\begin{table}[h]
\caption{
 Best-fit Schechter parameters of the composite luminosity function of five
 SDSS bands. The field values are from Blanton et al. (2001), whose
 paramters were shifted to match our cosmology.
 Galaxies within 0.75 Mpc are used. 
}\label{tab:5color}
\begin{center}
\begin{tabular}{lllll}
\hline
Band & $M^*$ & $\alpha$& Field  $M^*$ &  Field $\alpha$\\
\hline
\hline
$u$ & $-$21.61$\pm$0.26 & $-$1.40$\pm$0.11 & $-$19.11$\pm$0.08 & $-$1.35$\pm$0.09    \\
$g$ & $-$22.01$\pm$0.11 & $-$1.00$\pm$0.06 & $-$20.81$\pm$0.04 & $-$1.26$\pm$0.05    \\
$r$ & $-$22.21$\pm$0.05 & $-$0.85$\pm$0.03 & $-$21.60$\pm$0.03 & $-$1.20$\pm$0.03    \\
$i$ & $-$22.31$\pm$0.08 & $-$0.70$\pm$0.05 & $-$22.03$\pm$0.04 & $-$1.25$\pm$0.04    \\
$z$ & $-$22.36$\pm$0.06 & $-$0.58$\pm$0.04 & $-$22.32$\pm$0.05 & $-$1.24$\pm$0.05    \\
$r$(spec) & $-$22.31$\pm$0.13 & $-$0.88$\pm$0.07 & $\cdots$ & $\cdots$   \\
\hline
\end{tabular}
\end{center}
\end{table}

\begin{table}[h]
\caption{
\label{tab:dev_exp}
Best-fit Schechter parameters for de Vaucouleur and exponential galaxies in
 five SDSS bands. The galaxies are divided into two subsamples using profile
 fitting. Galaxies within 0.75 Mpc are used. 
}
\begin{center}
\begin{tabular}{lllll}
\hline
Band & $M^*$ (deV) & $\alpha$ (deV)&  $M^*$(exp) &  $\alpha$ (exp)\\
\hline
\hline
$u$ & $-$21.64$\pm$0.30 &  $-$1.41$\pm$0.12 &  $-$21.45$\pm$0.13 & $-$1.27$\pm$0.07    \\
$g$ & $-$21.92$\pm$0.11 & $-$0.73$\pm$0.07 & $-$21.89$\pm$0.13 & $-$1.20$\pm$0.06   \\
$r$ & $-$22.01$\pm$0.07 & $-$0.37$\pm$0.06 & $-$21.73$\pm$0.12 & $-$1.04$\pm$0.06     \\
$i$ & $-$22.13$\pm$0.07 & $-$0.25$\pm$0.06 & $-$21.69$\pm$0.13 & $-$0.80$\pm$0.08    \\
$z$ & $-$22.24$\pm$0.06 & +0.12$\pm$0.06 & $-$21.76$\pm$0.11 & $-$0.65$\pm$0.07    \\
\hline
\end{tabular}
\end{center}
\end{table}

\begin{table}[h]
\caption{
\label{tab:5color_cin}
Best-fit Schechter parameters for low concentration (early-type) and high
 concentration (late-type) galaxies in
 five SDSS bands.The concentration index here is  the ratio of 50\%
 Petrosian flux radius to 90\% Petrosian flux radius. 
 Early-type galaxies have a concentration of $<$0.4, and
 late-type galaxies have a concentration of $\geq$0.4.  Galaxies within 0.75 Mpc are used. 
}
\begin{center}
\begin{tabular}{lllll}
\hline
Band & $M^*$ (Early) & $\alpha$ (Early)&  $M^*$(Late) &  $\alpha$ (Late)\\
\hline
\hline
$u$ &  $-$21.42$\pm$0.24 & $-$1.28$\pm$0.12 & $-$21.82$\pm$0.11 & $-$1.42$\pm$0.06    \\
$g$ &  $-$22.05$\pm$0.11 & $-$0.89$\pm$0.07 & $-$22.26$\pm$0.11 & $-$1.36$\pm$0.05    \\
$r$ & $-$22.31$\pm$0.06 & $-$0.92$\pm$0.04 & $-$22.24$\pm$0.12 & $-$1.32$\pm$0.06     \\
$i$ & $-$21.97$\pm$0.09 & $-$0.59$\pm$0.10 &    $-$22.02$\pm$0.13 & $-$1.04$\pm$0.08    \\
$z$ & $-$22.08$\pm$0.09 & $-$0.47$\pm$0.09 &    $-$22.09$\pm$0.12 & $-$0.87$\pm$0.07    \\
\hline
\end{tabular}
\end{center}
\end{table}

\begin{table}[h]
\caption{
\label{tab:5color_ur}
Best-fit Schechter parameters for $u-r>$ 2.2 (early type) and $u-r\leq$
 2.2 (late type) galaxies in
 five SDSS bands. Galaxies within 0.75 Mpc are used. 
}
\begin{center}
\begin{tabular}{lllll}
\hline
Band & $M^*$ (Early) & $\alpha$ (Early)&  $M^*$(Late) &  $\alpha$ (Late)\\
\hline
\hline
$u$ &  $-$21.65$\pm$0.26 & $-$1.47$\pm$0.11 &  $-$21.78$\pm$0.13 &  $-$1.37$\pm$0.07    \\
$g$ & $-$22.04$\pm$0.10 & $-$1.03$\pm$0.06 & $-$22.30$\pm$0.09 & $-$1.38$\pm$0.05    \\
$r$ & $-$22.29$\pm$0.04 & $-$0.97$\pm$0.02 & $-$22.22$\pm$0.12 & $-$1.41$\pm$0.06     \\
$i$ & $-$21.91$\pm$0.08 & $-$0.58$\pm$0.07 & $-$22.17$\pm$0.16 &  $-$1.23$\pm$0.08    \\
$z$ & $-$21.93$\pm$0.07 & $-$0.36$\pm$0.08 & $-$22.14$\pm$0.19 & $-$1.08$\pm$0.09    \\
\hline
\end{tabular}
\end{center}
\end{table}

\begin{table}[h]
\caption{
\label{tab:bcg_center}
Best-fit Schechter parameters for galaxies using positions of brightest cluster
 galaxies as a  center in five SDSS bands. The mean deviation from the CE
 center used in this work is 1.02 arcmin.
}
\begin{center}
\begin{tabular}{lll}
\hline
Band & $M^*$  & $\alpha$ \\
\hline
\hline
$u$ & $-$21.84$\pm$0.16 & $-$1.43$\pm$0.07\\
$g$ & $-$22.16$\pm$0.15 & $-$1.05$\pm$0.07 \\
$r$ & $-$22.29$\pm$0.05 & $-$0.91$\pm$0.03 \\
$i$ & $-$22.31$\pm$0.06 & $-$0.73$\pm$0.03 \\
$z$ & $-$22.18$\pm$0.07 & $-$0.55$\pm$0.07 \\
\hline
\end{tabular}
\end{center}
\end{table}

\begin{table}[h]
\caption{
\label{tab:global}
 Best-fit Schechter parameters for galaxies using global background
 subtraction  in five SDSS bands. Instead of the annuli around the
 cluster, the global background was used to subtract the background
 galaxies to see the dependence on the background subtraction. 
}
\begin{center}
\begin{tabular}{lll}
\hline
Band & $M^*$  & $\alpha$ \\
\hline
\hline
$u$ &  $-$21.77$\pm$0.17 & $-$1.47$\pm$0.07 \\
$g$ &  $-$22.01$\pm$0.12 & $-$1.06$\pm$0.07 \\
$r$ &  $-$22.20$\pm$0.05 & $-$0.90$\pm$0.03 \\
$i$ &  $-$22.24$\pm$0.07 & $-$0.72$\pm$0.04 \\
$z$ &  $-$22.10$\pm$0.06 & $-$0.50$\pm$0.06 \\
\hline
\end{tabular}
\end{center}
\end{table}

\begin{table}[h]
\caption{
\label{tab:richness}
 Best-fit Schechter parameters in
 the $r$ band for galaxies using richer systems.
 The best-fit Schechter parameters for $N_{-18}>$20 and  
 $N_{-18}>$40  subsamples are shown. $N_{-18}$ is defined as
 the number of galaxies brighter than $-$18th magnitude after subtracting the background.
}
\begin{center}
\begin{tabular}{llll}
\hline
Band & $M^*$  & $\alpha$ & $N$(cluster) \\
\hline
\hline
$N_{-18}$ $>$20 &  $-$22.21$\pm$0.05 & $-$0.85$\pm$0.03 & 204  \\
$N_{-18}$ $>$40 &  $-$22.29$\pm$0.06 & $-$0.90$\pm$0.04 & 120  \\
\hline
\end{tabular}
\end{center}
\end{table}

\begin{table}[h]
\caption{\label{tab:previous}
 Comparison with previous studies on the composite luminosity function. The CE
 composite LFs (this work) was re-calculated using each
 author's cosmology. The magnitude was transformed using data from Fukugita et
 al. (1995) and Lumsden et al. (1992).
}
\begin{center}
\begin{tabular}{llllll}
\hline
Paper & $M^*$ & $\alpha$ & Band & Ncluster & Cosmology  \\
\hline
\hline 
CE   & 	$-$22.21$\pm$0.05  &   $-$0.85$\pm$0.03 & $r$ & 204 &
 $\Omega_{\rm{M}}$=0.3 $\Omega_{\Lambda}$=0.7 $H_0$=70 \\
\hline
Colless 89 &	$-$20.04 &	$-$1.21  &	$bj$ & 14
 rich & 	$H_0$=100 $q_0$=1	\\
(CE)  &	$-$21.58$\pm$0.12	& $-$0.93$\pm$0.06 & $bj$ & 204 &
 $H_0$=100 $q_0$=1	\\
 (CE) &	  $-$22.20$\pm$0.12	  &	$-$1.21 fixed  &	$bj$
 & 204 & $H_0$=100 $q_0$=1 	\\
\hline
Lugger 89 &	$-$22.81$\pm$0.13 	& $-$1.21$\pm$0.09  &  $R$ (PDS)   &  9 &	$H_0$=50	\\
(CE)  &	$-$22.49$\pm$0.06	& $-$0.69$\pm$ 0.05 & $R$ (PDS)  &  204 &
 $H_0$=50 $q_0$=0.5	\\
(CE)  &	$-$22.77$\pm$0.17 & $-$1.21 fixed & $R$ (PDS)  &  204 &	$H_0$=50 $q_0$=0.5	\\
\hline
Valotto 97 &	$-$20.0$\pm$0.1 	& $-$1.4$\pm$0.1  &  $bj$   &  55 Abell APM &	$H_0$=100 \\
(CE)  &	$-$21.58$\pm$0.12	& $-$0.93$\pm$0.06 & $bj$   &  204 & $H_0$=100 $q_0$=1	\\
(CE)  &	$-$22.69$\pm$0.23 & $-$1.4 fixed & $bj$  &  204 &	$H_0$=100 $q_0$=1	\\
\hline
Lumsden 97 &	$-$20.16$\pm$0.02  &	$-$1.22$\pm$0.04  &	$bj$
 & 22 rich & $H_0$=100 $q_0$=1 	\\
 (CE) &	  $-$21.58$\pm$0.12	  &	$-$0.93$\pm$0.06  &	$bj$	 & 204 &
 $H_0$=100 $q_0$=1 \\
 (CE) &	  $-$22.22$\pm$0.10	  &	$-$1.22 fixed  &	$bj$
 & 204 & $H_0$=100  $q_0$=1	\\
\hline
Garilli 99 & $-$22.16$\pm$0.15 & $-$0.95$\pm$0.07 & $r$ (CCD)  & 65 Abell
 X-ray & $H_0$=50 $q_0$=0.5 \\
 (CE) &	 $-$22.15$\pm$0.06 & $-$0.69$\pm$ 0.05 &	$r$ (CCD)   & 204 & $H_0$=50 $q_0$=0.5 \\
 (CE) &	 $-$22.28$\pm$0.05 & $-$0.84 fixed &	$r$ (CCD)   & 204 & $H_0$=50 $q_0$=0.5 \\
\hline
Paolillo 00 &	$-$22.26$\pm$0.16 	& $-$1.11  &  $r$ (POSSII)  &  39 Abell &	$H_0$=50 $q_0$=0.5	\\
(CE)  &	$-$22.15$\pm$0.06	& $-$0.69$\pm$ 0.05 & $r$ (POSSII)  &  204 &
 $H_0$=50 $q_0$=0.5	\\
(CE)  &	$-$22.55$\pm$0.12 & $-$1.11 fixed & $r$ (POSSII)  &  204 &	$H_0$=50 $q_0$=0.5	\\
\hline
Yagi 02 &	$-$21.3$\pm$0.2 	& $-$1.31$\pm$0.05  &  $R_C$  &  10 Abell &	$H_0$=100 $q_0$=0.5	\\
(CE)  &	$-$21.89$\pm$0.10 & $-$1.03$\pm$ 0.05 & $R_C$   &  204 &
 $H_0$=100 $q_0$=0.5	\\
(CE)  &	$-$22.55$\pm$0.14   & $-$1.31 fixed & $R_C$  &  204 &  $H_0$=100 $q_0$=0.5	\\
\hline
\end{tabular}
\end{center}
\end{table}

\clearpage

\end{document}